# Vortex motion in tilted magnetic fields in highly layered electron-doped superconductor $Nd_{2-x}Ce_xCuO_4$

O.E. Petukhova[1], A.S. Klepikova[1], M. R Popov[1], N.G. Shelushinina[1], A.A. Ivanov[2], T.B. Charikova[1]

[1]M.N. Mikheev Institute of Metal Physics, Ural Branch, Russian Academy of Sciences, 18, S. Kovalevskoy St., Ekaterinburg, 620108, Russia

[2]National Research Nuclear University MEPhI, Moscow, 115409, Russia

**Abstract**

The carrier transport and the motion of a vortex system in a mixed state of an electron-doped high-temperature superconductors $Nd_{2-x}Ce_xCuO_4$ were investigated. To study the anisotropy of galvanomagnetic effects of highly layered NdCeCuO system we have synthesized $Nd_{2-x}Ce_xCuO_4/SrTiO_3$ epitaxial films with non-standard orientations of the *c*-axis and conductive $CuO_2$ layers relative to the substrate. The variation of the angle of inclination of the magnetic field, ***B***, relative to the current, ***J***, reveals that the behavior of both the in-plane $\rho_{xx}(B)$ and the out-plane $\rho_{xy}(B)$ resistivities in the mixed state is mainly determined by $B_\perp$, the perpendicular to ***J*** component of ***B***, that indicates the crucial role of the Lorentz force, $\boldsymbol{F_L} \sim [\boldsymbol{J \times B}]$ and defines the motion of Josephson vortices across the $CuO_2$ layers.

## 1.Introduction

After the discovery of high-temperature superconductors (HTSC), assumptions were made that their high critical temperature is also due to their layered structure. The presence of conductive layers causes an appreciable anisotropy of the electromagnetic and superconducting properties of HTSC. One of the HTSC compounds with high anisotropy is electron-doped superconductor $Nd_{2-x}Ce_xCuO_4$.

The crystal structure of $Nd_{2-x}Ce_xCuO_4$ was described by Takagi et al. [1]. $Nd_{2-x}Ce_xCuO_4$ shows the tetragonal crystal structure (T' structure), which is characterized by the absence of apical oxygen atoms below and above the copper ions. The parent undoped compound $Nd_2CuO_4$ is an insulator. Also the parent compound exhibit another important characteristic: his ground state is antiferromagnetically ordered. With doping (substituting $Nd^{3+}$ with $Ce^{4+}$) the compound becomes superconducting for values of $x$ between 0.135 and 0.20 [2]. Furthermore, in the superconducting doped compound the long-range antiferromagnetic order of the Cu moments is destroyed, but the short-range antiferromagnetic fluctuations are presented up to $x \approx 0.15$ [3].

In the mixed state of conventional type II superconductors external magnetic field penetrates in a depth of type-II superconductor in the form of quantized vortices. These vortices in the superconducting layer distribute in a regular triangular lattice [4].

However, in highly layered materials such as $Nd_{2-x}Ce_xCuO_4$ the coherence length along the $c$-axis becomes shorter than the separation between the superconducting $CuO_2$ planes and vortex cores will be confined to the conduction planes, forming what is known as pancake vortices [5]. The coupling between pancake vortices in different layers occurs via Josephson interactions [6]. If the coupling is weak, superconducting system can be considered as a stack of decoupled two dimensional (2D) layers, otherwise the system can be regarded as anisotropic three-dimensional (3D) [7].

However, even in the case of an anisotropic 3D superconductor in a tilted magnetic field, Josephson and pancake vortices coexist and form a crossing lattice in a highly anisotropic superconductor with the attractive interaction between the two different kinds of vortices [8]. The effect of the crossing lattices of Abrikosov and Josephson vortices is not well understood to date.

In this work, we analyzed carrier transport and the motion of a vortex system in the electron-doped HTSC $Nd_{2-x}Ce_xCuO_{4+\delta}$ in underdoped ($x = 0.145$) region, in the area of the evolution from antiferromagnetic (AFM) to superconducting (SC) order.

**2. Materials and methods**

For this purpose, we have synthesized by pulsed laser deposition $Nd_{2-x}Ce_xCuO_{4+\delta}/SrTiO_3$ epitaxial film ($1\bar{1}0$) with optimal oxygen content $\delta=0$ and $x = 0.145$ [9], where the $c$-axis of the $Nd_{2-x}Ce_xCuO_{4+\delta}$ lattice is directed along the short side of the $SrTiO_3$ substrate (Fig. 1).

In the process of pulsed laser deposition, an excimer KrF laser was used with a wavelength of 248 nm, with an energy of 80 mJ/pulse. The energy density at the target surface is 1.5 J/cm$^2$. The pulse duration was 15 ns, the repetition rate of pulses was from 5 to 20 Hz. Further, the synthesized film was subjected to heat treatment (annealing) under various conditions to obtain samples with a maximum superconducting transition temperature. X-ray diffraction analysis (Co-K radiation) showed that film was of high quality and was single crystal.

The optimum annealing conditions was $T^c_{onset}$ = 15.7 K, $T_c$ = 10.7 K, $t$ = 60 min., $T$ = 600$^0$C, $p$ = 10$^{-5}$Torr. The thickness of the film was $d$ = 520 nm.

For correct measurement using the standard four-probe method, the geometry of the samples was six-contact Hall bar. In the process of the Hall effect measure in superconductors, there are several measurement errors, which in magnitude can be larger than the measured signal:

a. Misalignment voltage connected with imperfect sample geometry. These voltages are not magnetic field dependent and can be compensated by use field reversal to remove by subtraction.

b. Thermal electric voltage due to temperature effects. These voltages are not current dependent and can be compensated by use current reversal to remove by subtraction.

Thus, four measurements (for two directions of current and voltage) were performed to separate the Hall voltage from the misalignment voltage and thermal electric voltage.

The magnetic field dependences of the longitudinal, $\rho_{xx}(B,T)$, and Hall, $\rho_{xy}(B,T)$, resistivity for Nd$_{2-x}$Ce$_x$CuO$_{4+\delta}$/SrTiO$_3$ film were investigated in the Quantum Design PPMS 9 (Center for Nanotechnologies and Advanced Materials, IFM UrB RAS). The electric field was applied parallel to the SrTiO$_3$ substrate plane. The external magnetic field $B$ was always perpendicular to the $c$-axis (being parallel to the $ab$-planes) while it was directed at different angles to the supercurrent.

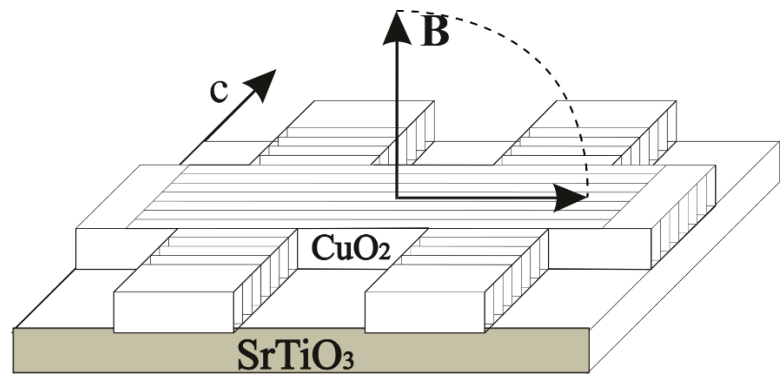

Figure 1. The orientations of conductive $CuO_2$ planes relative to the substrate $SrTiO_3$ for the $Nd_{2-x}Ce_xCuO_{4+\delta}$ film.

Transport measurements were performed for $Nd_{2-x}Ce_xCuO_{4+\delta}$ sample with a non-standard arrangement of $CuO_2$ layers - the layers were perpendicular to the substrate plane, with the *c*-axis located along the short side of the sample. In addition, the current was applied perpendicular to the *c*-axis along the long side of the sample.

This arrangement of the planes led to the fact that when a magnetic field was applied perpendicularly to the substrate, a lattice of Josephson vortices appeared along the field, while Abrikosov vortices (pancakes) were absent (Fig. 2). In our measurements, we changed the slope of the magnetic field from $\varphi = 90$ to $\varphi = 0$ degrees in such a way that the rotation axis was parallel to the *c*-axis. In this case, when the magnetic field was tilted, an additional field component appeared along the direction of the current, which at $\varphi = 0$ led to the appearance of Josephson vortices along the direction of the current.

All Josephson vortices are aligned parallel to the superconducting planes. The interaction with the inhomogeneous layered environment gives rise to the intrinsic pinning: the vortex energy becomes dependent on the vortex position with respect to the planes and produces a pinning force which tries to keep vortices in between the superconducting layers.

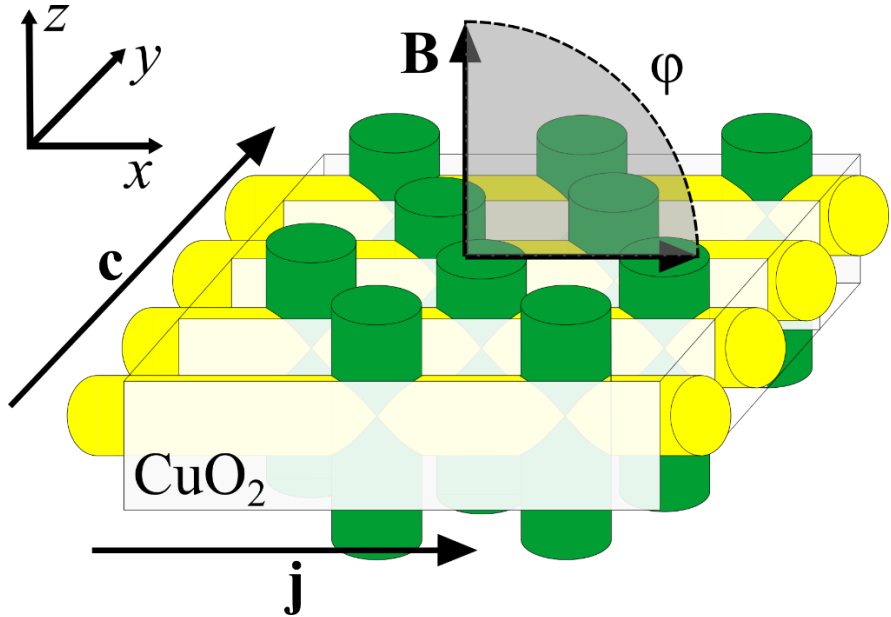

Figure 2. Schematic structure of vortexes orientation with the following configuration of electric current and magnetic field: the electric current (j) and the magnetic field (B) are parallel to $CuO_2$ planes; the angle $\varphi$ between electric current and magnetic field changes from $\varphi = 90$ to $\varphi = 0$ degrees. The green cylinders are the Josephson vortices for the perpendicular component of the magnetic field, the yellow cylinders are Josephson vortices for the parallel component of the magnetic field.

**3. Theoretical concepts**

*Layered superconductors.* The new oxide superconductors are layered compounds with building blocks made out of conducting (metallic) $CuO_2$ planes separated by buffer layers which serve as a charge reservoir. (see monographs [4], [5], [10], [11] for a detailed description). The transport properties are roughly uniaxial, with a large anisotropy between the *c* axis and the *ab* planes due to the layered structure and essentially isotropic behavior within the $CuO_2$ planes.

For not too large anisotropy, a description in terms of a continuous anisotropic Ginzburg-Landau or London theory is applicable. On the other hand, for the highly anisotropic high-*Tc* superconductors the discreteness of the structure becomes relevant, and they may be considered as a set of quasi-two-dimensional superconducting systems with weak Josephson connection between the layers [7], [12], [13], [14]. Such a description is provided by the discrete Lawrence-Doniach [7] model, which will provide the basis for the discussion of the physics of layered superconductors.

The criterion usually adopted to go from a continuous anisotropic to a discrete layered description is the smallness of the coherence length along the *c* axis, $\xi_c$, with respect to the layer separation *d* as expressed by the dimensionless ratio $\eta = 2\xi_c^2(0)/d^2$ (see, for example, [14]).

The ratio characterizes the crossover from quasi-2D layered to continuous 3D anisotropic behavior: for a large coherence length $\xi_c(0)$, i.e., $\eta \gg 1$, the continuous description is always appropriate; for small $\eta \ll 1$, a crossover will take place at a temperature

$$T_{c0} = (1 - \eta)T_c < T_c \qquad (1)$$

where the system behaves in a quasi-two-dimensional manner at low temperatures, $T < T_{c0}$, and exhibits 3D anisotropic behavior above $T_{c0}$.

*Josephson vortices.* In a layered superconductor, the structure of the individual vortices as well as that of the vortex lattice can be strongly modified (see, for example, review [14]). Thus, for magnetic field directed perpendicular to the *c* axis, an individual vortex line is not a rectilinear object as the usual continuous Abrikosov vortex. Single flux-line in a strongly layered superconductor can be viewed as an array of pancake vortices threading the individual superconducting layers and interconnected by Josephson strings.

As already said, layered superconductors consist of a stack of alternating thin superconducting layers separated by non-superconducting regions of width $d$. For $d \gg \xi_c$ the superconducting layers are essentially two-dimensional (2D) as long as they are so thin that there is no variation in fields or in the superconducting order parameter across each layer. Such a superconductor can carry super-currents along the layers, as well as between the layers. This is due to the Josephson tunneling of Cooper pairs [15] across the insulating regions that separate neighboring superconducting layers, i.e., each pair of neighboring layers forms one Josephson junction.

We focus on the properties of the Josephson vortex lattice generated by the magnetic field applied along the layers. The theoretical description is based on the Lawrence-Doniach model [7]. In that case the Josephson character of interlayer coupling becomes most pronounced because interlayer currents are involved in the formation of vortices.

At low fields the Josephson vortices are isolated and form a triangular lattice, strongly stretched along the layers. Josephson vortices do not have normal cores, in contrast with Abrikosov vortices in standard anisotropic superconductors. Rather, the centers of these vortices lie between layers, and the normal core is replaced by the nonlinear phase core, the region within which the phase difference between the two layers sweeps from 0 to $2\pi$ [12], [13], [14], [16].

The core size is given by the Josephson length $\lambda_J = \gamma d$ along the *ab* plane (i.e., along layers) and length *d* along the *c* axis (where the Josephson character of the interlayer current is important). Here $\gamma$ is the anisotropy of the London penetration depth, $\gamma = \lambda_c/\lambda_{ab}$ where $\lambda_c$ and $\lambda_{ab}$ are the penetration depths for currents perpendicular and parallel to the layers, respectively.

When magnetic field exceeds the crossover field

$$B_{cr} = \Phi_0/\pi\gamma d^2, \qquad (2)$$

the cores of Josephson vortices start to overlap and a dense Josephson lattice begins to form [13], [16], [17]. In fields $B \gg B_{cr}$ all interlayer spacings are filled by vortices and the vortex lattice in a given interlayer spacing becomes very similar to that in a single Josephson junction.

*Intrinsic pinning.* Intrinsic pinning of Josephson vortices in the oxides is a consequence of the layered structure of the material, with strong superconductivity present in the metallic $CuO_2$ planes and only weak or vanishing superconductivity found in the intermediate buffer layers. Therefore, the superconducting order parameter (and, together with it, the condensation energy) is expected to exhibit strong oscillations with period $d$, the interlayer distance(see for details [4], [14]).

The first quantitative analysis of intrinsic pinning in layered superconductors was that of Tachiki and Takahashi [18]. They proposed that the modulation of the order parameter perpendicular to the layers can pin the vortices between the layers thus leading to an intrinsic pinning mechanism of the Josephson vortices when the magnetic field is parallel to the $CuO_2$ planes.

With the magnetic-field $B$ aligned parallel to the $ab$ planes ($y$ axis), the vortex lattice tries to accommodate itself to the layer structure so that the vortex cores come to lie in between the strongly superconducting $CuO_2$ planes. A current density **j** ($\parallel x$ axis) flowing along the planes will exert a Lorentz force on the vortices which is pointing along the $c$ axis ($z$ axis of the coordinate system).

In order to move, the vortices have to cross the strongly superconducting layers, which involves a large expenditure of condensation energy, thus creating the intrinsic pinning barriers. It is this special geometry in which Josephson vortex tunneling takes place.

### 4. Experimental results and discussion

The cerium-doped cuprate $Nd_{2-x}Ce_xCuO_{4+\delta}$ with an electronic conductivity type has a layered quasi-two-dimensional perovskite-like crystal structure with a body-centered crystal lattice and corresponds to a tetragonal T'-phase. The structure contains a single $CuO_2$ plane per unit cell with the distance between adjacent $CuO_2$ planes, $d = 0.6$ nm [19]. Lattice parameters are $a = b = 0.394$ nm, $c = 1.208$ nm. Optimally annealed ($\delta \rightarrow 0$) $Nd_{2-x}Ce_xCuO_{4+\delta}$ crystals have clearly pronounced two-dimensional properties [20].

According our investigation we have found that in the framework of the model of a natural superlattice [10], [21], [22], [23], [24] in which $CuO_2$ layers represent quantum wells and the buffer layers serve as barriers two complementary processes determine conductivity along the $c$-axis: incoherent tunneling and thermal activation across the barriers [25] for optimally annealed compounds $Nd_{2-x}Ce_xCuO_4$ with $x = 0.12 – 0.15$. The conductivity in $CuO_2$ layers have a metallic-like behavior and resistivity $\rho_{ab} \sim T^2$ in a wide temperature range.

The condition (1) for the temperature crossover from 3D to quasi-2D behavior of a layered superconductor can also be represented as [7], [26]:

$$\gamma^2(1 - t_{c0}) = 2\left(\frac{\xi_{ab}(0)}{d}\right)^2, \qquad (3)$$

where $t_{c0} = T_{c0}/T_c$ is the crossover temperature, $\xi_{ab}$ is the Ginzburg-Landau coherence length in $CuO_2$ layers and the anisotropy coefficient, $\gamma = \lambda_c/\lambda_{ab}$, can be expressed as follows

$$\gamma^2 = \frac{m_c}{m_{a,b}} = \frac{\rho_c}{\rho_{ab}} = \left(\frac{\xi_{ab}(0)}{\xi_c(0)}\right)^2, \qquad (4)$$

where $m_c$ denotes the effective mass along the c-axis, $m_{a,b}$ is the corresponding planar parameter.

For $Nd_{2-x}Ce_xCuO_4$ with $x = 0.145$ we obtained $\gamma^2 = 1600$, $\xi_{ab}(0) = 110$ Å [9] and for $d = 6$ Å we have got $1-t_{c0} \sim 0.42$. Considering that $T_c = 10.7$ K for $Nd_{2-x}Ce_xCuO_4$ with $x = 0.145$ we obtained the 3D - 2D crossover temperature is equal $T_{c0} = 6.2$ K. So, the compound $Nd_{2-x}Ce_xCuO_4$ with $x = 0.145$ will show 2D behavior below $T_{c0} = 6.2$ K. On the other hand the compound $Nd_{2-x}Ce_xCuO_4$ with $x = 0.135$ ($\gamma^2 \sim 800$, $\xi_{ab}(0) = 122.6$ Å, $T_c = 9.6$ K) always behave like an anisotropic 3D superconductor.

In-plane magnetoresistivity $\rho_{xx}^{ab}(B)$ (Fig. 3a) and magnetic field dependencies of the out-of-plane Hall resistivity $\rho_{xy}(B)$ (Fig. 3b) for $Nd_{2-x}Ce_xCuO_4/SrTiO_3$ films with doping level $x = 0.145$ at the temperature $T = 1.8$ K and transport current $I = 50$ μA were investigated at different angles between the magnetic field and the direction of the transport current.

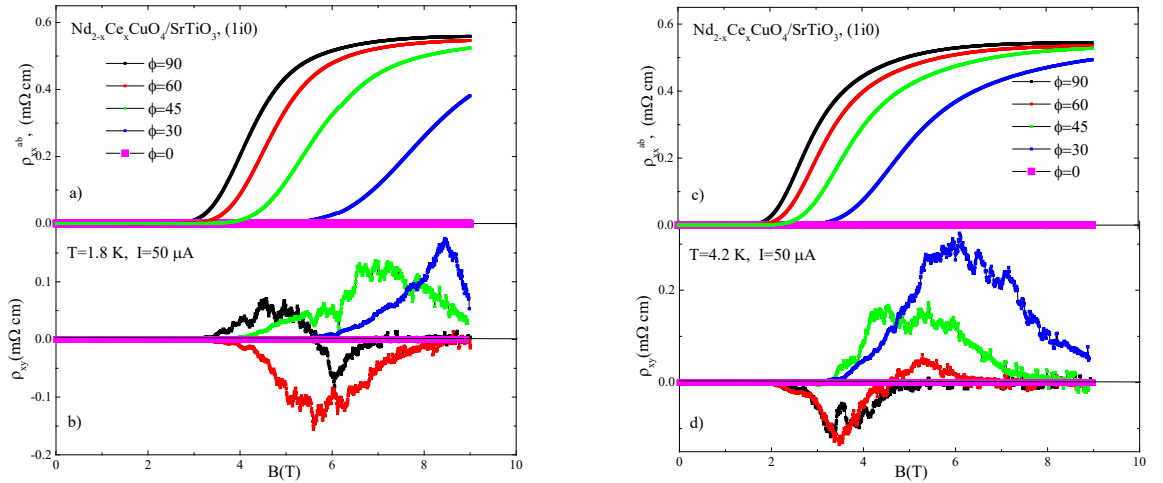

Figure 3. The magnetic field dependencies of the in-plane magnetoresistivity $\rho_{xx}^{ab}(B)$ at $T=1.8$ K (a) and 4.2K (c) and out-of-plane Hall resistivity $\rho_{xy}^{c}(B)$ at $T=1.8$ K (b) and 4.2K (d) for $Nd_{2-x}Ce_xCuO_4/SrTiO_3$ films with $x=0.145$ and optimal annealing for transport current I = 50 μA at different tilt of the magnetic field relative to the direction of the current.

As we can see at $\varphi = 90^0$ there are three regions in the magnetic fields of different states of the system: the normal state in strong magnetic field B ≅ (6 – 9) T with the in-plane resistivity $\rho_{xx}^{ab}$ and $\rho_{xy}^{c} \cong 0$, the region of the mixed state B ≅ (3 –6) T and the region of the superconductivity in weak magnetic field B ≅ (0 – 3) T. The region of the mixed state demonstrates the smooth decrease of the in-plane magneto-resistance and non-trivial double sign change in magnetic field dependence of the out-of-plane Hall resistivity. In the superconducting region in the magnetic field B = (0 – 3) T both of the in-plane magnetoresistivity and out-of-plane Hall resistivity are equal zero.

The decrease of the angle between the direction of the magnetic field and the direction of the transport current leads to the superconducting transition broadening and to the shift to the region of higher magnetic field. In the situation when the magnetic field is co-directed with the direction of the transport current, the magnetic field B = 9 T is not enough to destroy superconductivity.

The out-of-plane Hall resistivity $\rho_{xy}^{c}(B)$ shows an unusual peak value in the mixed region: at small angles between the direction of the magnetic field and the direction of the transport current ($\varphi = 30^0$, $45^0$) this peak is positive, at large angle $\varphi = 60^0$ it is negative and at $\varphi = 90^0$ it has double sign change form.

The similar in-plane magnetoresistivity $\rho_{xx}^{ab}(B)$ (Fig. 3c) and magnetic field dependencies of the out-of-plane Hall resistivity $\rho_{xy}(B)$ (Fig.3d) at different angles between the direction of the magnetic field and the direction of the transport current for $Nd_{2-x}Ce_xCuO_4/SrTiO_3$ films with doping level $x= 0.145$ and transport current I = 50 μA were found at the temperature $T = 4.2$K. As we can see, the resistivity value and the general course of the dependencies $\rho_{xx}^{ab}(B)$ and $\rho_{xy}^{c}(B)$ did not change, however, the beginning of the superconducting transition shifted to the region of lower fields.

Let us first discuss the $\rho_{xx}^{ab}(B)$ dependences. Figs 3a and 3c show the results of $\rho_{xx}(B)$ for the sample, the geometry of which is shown in Fig. 1 and 2. It is in this geometry that the often discussed "intrinsic pinning" for Josephson vortices is realized. The $\rho_{xx}(B)$ dependences have the form typical for type II superconductors under an external current along the *x* axis and a magnetic field perpendicular to it. Magnetic field applied along the layers creates the lattice of Josephson vortices and transport properties of a mixed state are determined by dynamics of the Josephson lattice [26], [27], [28].

The vortex lines in type II superconductors are subject to the Lorentz force and begin to flow perpendicular to the current and magnetic field when the Lorentz force exceeds the pinning force at $B = B_{dp}$ (depinning field). This is the flux flow regime where the motion of the vortex lattice results in energy dissipation and a finite resistivity arises [5].

Let's estimate the critical field, $B_{cr}$, where the cores of Josephson vortices start to overlap. According to (2), for the sample under study ($\gamma = 40$, $d = 0.6$ nm) we find $B_{cr} = 45.8\ T$ and the field region $B < 9T$ corresponds to the regime of a dilute Josephson vortex lattice.

As far as we know the theoretical calculations of the flux flow resistivity for the Josephson vortices are made only for the following geometric situation: the current ***J*** is directed along the c-axis, ***B***⊥ the c-axis, ***B*** ⊥ ***J*** [27], [28]. Note that a current density running along the *c*- axis will push the vortices to move along the planes (*x* axis) where no barriers are inhibiting the flow. In this case, point defects are essential for producing a pinning.

Clem and Coffey (CC) [28] take into account the discreteness of the copper oxide planes when the coherence length $\xi_c(T)$ becomes less than the lattice constant *d*. The CC model yields the linear field dependence of the flux flow resistivity

$$\rho_f = 2.8\ (m_c/m_b)^{1/2} \frac{B}{(\Phi_0/d^2)} \rho_c \qquad (5)$$

for current in the *c* direction, magnetic field along the *a* direction and vortex motion in the *b* direction. For comparison, the anisotropic BS model [29] at that case gives $\rho_f = \rho_c(B/B_{c2a})$ where $B_{c2a} = \Phi_0/2\pi\xi_b\xi_c$ is the second critical field along the *a* axis.

In the BS model the dissipation that contributes to the viscous drag is concentrated in the vicinity of the normal vortex core and the anisotropic BS result for the flux flow resistivity contains a cross-sectional area of the vortex core region proportional to the product $\xi_b\xi_c$. In contrast, the CC model result contains an area proportional to $d^2(m_c/m_b)^{1/2} \sim d^2\lambda_c/\lambda_b$ owing to the atomic scale discreteness of the high-Tc superconductor.

The theoretical conceptions for dynamics of Josephson vortices in a layered superconductor are further developed by Koshelev [27] which showed that the field behavior of $\rho_f$ depends on the mechanism of dissipation. Moving Josephson vortices generate both in-plane and inter-plane electric fields, which induce dissipative quasiparticle currents. In [28] only dissipation due to the tunneling of quasiparticles between the layers was taken into account in calculation of the viscosity coefficient.

However, in the high-Tc superconductors, the in-plane quasiparticle conductivity $\sigma_{ab}$ is strongly enhanced in superconducting state as compared to the normal conductivity while the *c*-axis component $\sigma_c$ rapidly decreases with temperature in superconducting state. Below the transition temperature the anisotropy of dissipation $\sigma_{ab}/\sigma_c$ becomes larger than the superconducting anisotropy $\gamma^2$. This leads to dominating role of the in-plane dissipation in dynamics of the Josephson lattice. A field dependence of the flux-flow resistivity $\rho_f(B)$ then obtains a form:

$$\rho_f = \frac{B^2}{B^2+B_\sigma^2}\rho_c \tag{6}$$

with $\rho_c$ being the flux-flow saturation resistivity along the *c*-axis, $B_\sigma = (\frac{\sigma_{ab}}{\sigma_c})^{1/2} \Phi_0/\sqrt{2\pi}\, \gamma^2 d^2$.

Thus, for strong in-plane dissipation the indicated dependence $\rho_f(B)$ should have pronounced upward curvature at $B < B_\sigma$ and approaches $\rho_c$ at $B >> B_\sigma$, as shown schematically in Fig. 1 of Ref. 3. The $\rho_c(B)$ dependence of this type in a mixed state of a highly layered electron-doped superconductor $Nd_{2-x}Ce_xCuO_4$ (*x*=0.135, 0.145 and 0.15) was experimentally observed in our work [30] for films geometrically corresponding to calculations of Ref. 3.

Figs 4 show our dependencies $\rho_{xx}^{ab}(B_\perp)$ for different $\phi$ of the magnetic field at T = 1.8K and 4.2K as well as the fitting of the dependence (5) to the experimental curves at T = 1.8K. It can be seen from the figures that the experimental curves for our geometry are empirically well described by a formula of the type (5) with $B \rightarrow B_\perp = B\sin\phi$. This possibly indicates the important role of the in-plane dissipation channel, although, of course, specific theoretical calculations of the flux-flow resistivity are required for the situation with intrinsic pinning of Josephson vortices.

Figures 4 also show that in a case studied an almost parallel shift of the curves $\rho_{xx}(B)$ with the temperature is observed, in contrast to the case with Abrikosov vortices, when a broadening of these dependences with *T* is usually observed.

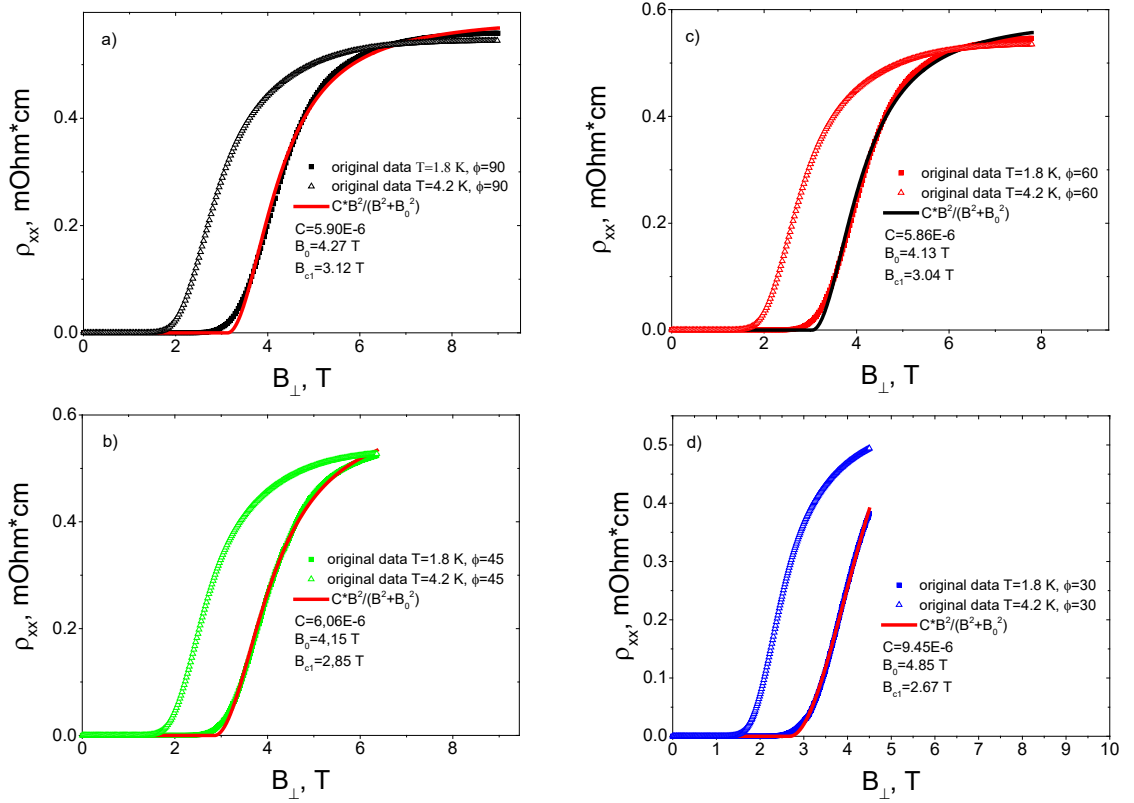

Figure 4(a-d). The dependencies $\rho_{xx}^{ab}(B)$ for different tilt angles of the magnetic field at T =1.8K and 4.2K. A fitting of the dependence (5) to the experimental curves is shown for T =1.8K.

One thing is certain, namely, the decisive role of the Lorentz force

$$\boldsymbol{F_L} \propto [\boldsymbol{J} \times \boldsymbol{B}], \qquad (7)$$

leading to the motion of Josephson vortices and, as a consequence, to dissipation processes. This is clearly seen if we present the results as a function of $B_\perp = B\sin\phi$ (see Figs 5a-d). Comparison of the $\rho_{xx}(B_\perp)$ curves in Figs 4 for different values of $\phi$ at a fixed $T$ also confirms the dependence of the observed resistivity only on the magnetic field component perpendicular to the external current.

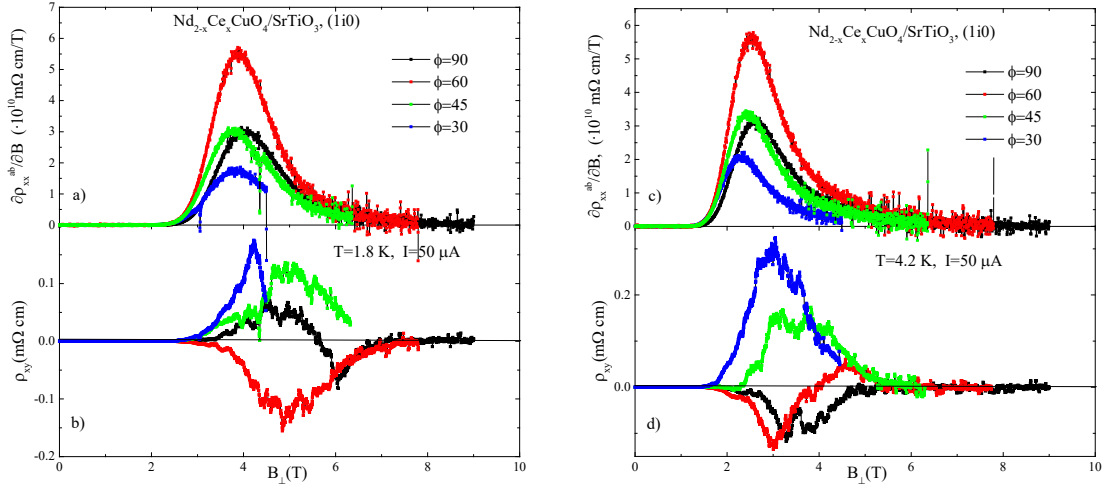

Figure 5. The dependencies of the in-plane magnetoresistance magnetic field derivative, $\partial\rho_{xx}^{ab}/\partial B$, (a,c), and of out-of-plane Hall resistivity, $\rho_{xy}^{c}$, (b,d), on the perpendicular to current magnetic field component, $B_{\perp} = B\sin\phi$, for $Nd_{2-x}Ce_xCuO_4/SrTiO_3$ films with $x$=0.145 and optimal annealing at $T$=1.8 K (a,b) and $T$ = 4.2 K (c, d) for transport current I = 50 μA at different tilt of the magnetic field.

Another interesting feature of layered superconductors is the unusual behavior of the upper critical magnetic field. For orientation parallel to the layers, it grows rapidly as temperature decreases and $\xi_c(T)$ approaches the interlayer distance $d$ (see, for example, [4] and references therein).

In the continuous anisotropic case the upper critical field is known to be [7]:

$$B_{c2} = \frac{\Phi_0}{2\pi\xi_{ab}\xi_c} \qquad (8)$$

The highly layered case, which realized for the relation $\xi_c(0) > d$ between the interlayer distance $d$ and the coherence length $\xi_c$, corresponds to a high upper critical field

$$B_{c2} = \frac{\Phi_0}{2\pi d^2}\frac{\xi_c}{\xi_{ab}}\frac{1}{\sqrt{1-(d^2/2\xi_c^2)}} \qquad (9)$$

We observe that the upper critical field diverges as the temperature-dependent coherence length $\xi_c(T)$ approaches $d$ from above with lowering the temperature [31]. In this limit, vortex cores fit in-between the superconducting layers, and the supercurrents do not destroy superconductivity.

Bulaevskii [32] was the first to show that an inhomogeneous state should exist in highly layered superconductors in parallel fields, and the field of transition from this state to the normal state at zero temperature is determined by the paramagnetic effect and equal to

$$B_p(0) = \sqrt{2}\Delta(0)/g\mu_B \qquad (10)$$

where $\Delta(0)$ is the value of the gap at $T = 0$, $\mu_B$ is the Bohr magneton, g being the Lande factor. For the investigated sample, the estimate according to (10) gives $B_p(0) >\approx 100T$ (for g = 2, $T_c = 15K$ [4], $\Delta(0)/kT_c = 3.5$). Indeed, it is seen from Fig. 3 that at $\varphi = 0$ the destruction of superconductivity and the transition to the normal state does not occur at least up to 9T.

For $\varphi \neq 0$, there is a component of the magnetic field perpendicular to the current, but still parallel to the CuO2 planes, so that the destruction of superconductivity at $B_\perp >\approx 6T$ (see Fig.5) is associated not with the achievement of a critical field, $B_{c2}$, but with the process of dissipation due to the motion of Josephson vortices perpendicular both to the current and the field under the action of the Lorentz force.

Let us now turn to the experimental data on the out-of-plane Hall resistivity $\rho_{xy}^c$ (B) (see Figs 3b, d and Figs 5b, d). We note that in the normal state there is no linear dependence of $\rho_{xy}$ on $B$, which is inherent in the conventional Hall resistance in a continuous medium, and also that the value of $|\rho_{xy}|$ is close to zero.

The same situation is observed for x = 0.135 and x = 0.15 $Nd_{2-x}Ce_xCuO_4$ films (see Fig. 2 in [30]) with a similar geometry corresponding to the motion of carriers (until a stationary state is established for the Hall voltage $V_y$) across the CuO2 layers under the action of the Lorentz force (cyclotron "twisting" of current carriers). This can be considered as a "lock" on cyclotron twisting for the out-of-plane motion of carriers due to the discreteness of a highly layered system in the 2D limit.

The $\rho_{xy}^c$ (B) behavior in a mixed state is much more rich that, logically, should be due to the motion of the vortex subsystem in the transverse direction under the action of the Lorentz force. However, as has been discussed repeatedly for the Abrikosov vortices (see [4] and references therein), the contribution to the Hall voltage from the lateral motion of the vortex subsystem is provided only if the charge of the vortex core differs (by an amount $\rho(0) = e\delta N$, $\delta N$ being the extra electron density) from the average charge of environment: $V_y \sim \delta N$.

Thus, Khomskii and Freimuth [33] and Feigel'man et al. [34], [35] proposed possible mechanisms for Abrikosov vortex charging by regarding the core region as the normal state and considering its chemical-potential difference from the environment. The resulting charge accumulation should decrease monotonically as $B$ increased because of the decreasing pair potential.

A somewhat different approach has been developed in a series of works by Kohno et al. [36], [37], [38] who explored the physics of the Lorentz force in superconductors. They studied Abrikosov vortex charging caused by the Lorentz force on supercurrent based on the quasiclassical equations of superconductivity. The study of vortex lattice in the range $B_{c1} < B < B_{c2}$ reveals that each
vortex core with a single flux quantum also accumulates charge due to the circulating supercurrent and has a Hall voltage across the core.

The field dependence of the charge density at the core center in their theory is well described by the non-monotonic field dependence:

$$\rho(0) \sim B(B_{c2} - B) \qquad (11)$$

with a peak near $B_{c2}/2$ originating from competition between the increasing magnetic field and the decreasing pair potential: $\rho(0) \sim B\Delta_{max}^2$. Here $\Delta_{max}$ denotes the maximum value of the pair potential in the vortex lattice, which, approaching $B_{c2}$, decreases as [39]

$$\Delta_{max} \sim (B_{c2} - B)^{1/2}. \qquad (12)$$

We observed the non-monotonic $\rho_{xy}(B)$ in a mixed state, which is qualitatively consistent with the results of the theory [36] - [38] ( cf. our Figs 3b,d with the graphs of Fig. 3 in the work [37]). We emphasize the important fact that the effect as a whole is due to the component of magnetic field perpendicular to the current, $B_\perp$ (see Figs 5b,d), which supports the hypothesis on the significant role of a Lorentz force in the vortex charge accumulation.

Although, in general, the $\rho_{xy}(B)$ dependences are within the scope of a relationship between $\rho_{xy}$ and $B_\perp$, the specific details of this behavior, up to a change in the sign of the effect, vary significantly with the variations in both the tilt angle of the magnetic field and the temperature. We believe that this may be due to some inhomogeneities in the *ab* planes, as well as to fluctuations (including thermal) of the tunnel probability of Josephson vortices moving across the planes under the conditions of intrinsic pinning.

## 5. Conclusions

This article is devoted to galvanomagnetic effects ($\rho_{xx}$, $\rho_{xy}$) in a mixed state of an electron-doped superconductor NdCeCuO, associated with the motion of Josephson vortices (the magnetic field is perpendicular to the $c$ - axis and parallel to the $ab$ planes).

Due to the advances in the growing technology of the high-quality $Nd_{2-x}Ce_xCuO_4/SrTiO_3$ epitaxial films with the different orientations of the $c$-axis relative to the substrate we have managed to realize, in a magnetic field parallel to the $CuO_2$ layers, the situation of Josephson vortices localized by an intrinsic pinning.

The variation of the tilt angle of the magnetic field, **B,** relative to the current, **J,** reveals that the behavior of both the in-plane $\rho_{xx}(B)$ and the out-plane $\rho_{xy}(B)$ resistivities in the mixed state is mainly determined by $B_\perp$, the perpendicular to **J** component of **B.** This fact clearly indicates the crucial role of the Lorentz force, $\boldsymbol{F_L} \sim [\boldsymbol{J} \times \boldsymbol{B}]$, that defines the motion of Josephson vortices across the $CuO_2$ layers.

Our research reveals that theoretical developments to describe the studied experimental situation on the Josephson vortices with intrinsic pinning are insufficient (for $\rho_{xx}$) or even completely absent (for $\rho_{xy}$). To interpret the data on $\rho_{xy}^c(B)$ we use some theoretical concepts developed for the Abrikosov vortices [36] - [38] that still requires a further analysis.

**Acknowledgments**

The study was supported by the Ministry of Science and Higher Education of the Russian Federation, agreement no. 075-15-2020-797 (13.1902.21.0024).